\documentclass[paper,twocolumn,twoside]{geophysics}
\usepackage{amsmath,amsfonts}
\usepackage{algorithmic}
\usepackage{algorithm}
\usepackage{array}
\usepackage{booktabs}
\usepackage{multirow}
\usepackage{multicol}
\usepackage{tabularx}
\usepackage{derivative}
\usepackage{hyperref}
\usepackage[english]{babel}
\usepackage{microtype}
\usepackage{enumitem}
\usepackage{graphicx}
\usepackage{soul}
\usepackage{xcolor}
\DeclareMathOperator*{\argminB}{argmin}   

\usepackage[switch]{lineno}

\begin{document}

\title{Poststack Seismic Data Preconditioning via Dynamic Guided Learning}

\renewcommand{\thefootnote}{\fnsymbol{footnote}} 

\address{

\footnotemark[1]Universidad Industrial de Santander, Department of Computer Science, Colombia. E-mail: javier.torres.quintero@hotmail.com, goyes.yesid@gmail.com,  luis2238326@correo.uis.edu.co, henarfu@uis.edu.co. 
\footnotemark[2]Universidad Industrial de Santander, Department of Physics, Colombia. E-mail: ana.mantilla@correo.uis.edu.co, jsanabri@uis.edu.co.      
}
\author{Javier Torres-Quintero\footnotemark[1], Paul Goyes-Pe\~nafiel\footnotemark[1], Ana Mantilla-Dulcey\footnotemark[2], Luis Rodríguez-López\footnotemark[1], José Sanabria-Gómez\footnotemark[2], Henry Arguello\footnotemark[1]}

\footer{Example}
\lefthead{Torres-Quintero et al.}
\righthead{Preconditioning via Dynamic Guided Learning}


\begin{abstract}
\
 Seismic data preconditioning is essential for subsurface interpretation. It enhances signal quality while attenuating noise, improving the accuracy of geophysical tasks that would otherwise be biased by noise. Although classical poststack seismic data enhancement methods can effectively reduce noise, they rely on predefined statistical distributions, which often fail to capture the complexity of seismic noise. On the other hand, deep learning methods offer an alternative but require large and diverse data sets. Typically, static databases are used for training, introducing domain bias, and limiting adaptability to new noise poststack patterns. This work proposes a novel two-process dynamic training method to overcome these limitations. Our method uses a dynamic database that continuously generates clean and noisy patches during training to guide the learning of a supervised enhancement network. This dynamic-guided learning workflow significantly improves generalization by introducing variability into the training data. In addition, we employ a domain adaptation via a neural style transfer strategy to address the potential challenge of encountering unknown noise domains caused by specific geological configurations. Experimental results demonstrate that our method outperforms state-of-the-art solutions on both synthetic and field data, within and outside the training domain, eliminating reliance on known statistical distributions and enhancing adaptability across diverse data sets of poststack data.
\end{abstract}

\section{Introduction}\

Seismic data preconditioning is crucial in preparing poststack seismic data for geophysical downstream tasks, particularly subsurface interpretation \citep{Chopra2013}. It enables geologists to analyze subsurface structures with greater accuracy in tasks such as fault detection \citep{WEI2022104968} and seismic attribute computation \citep{Oumarou2021}, facilitating the identification of promising areas for the exploration of mineral deposits, hydrocarbons, and geothermal resources \citep{gao2020threedimensionalseismiccharacterizationimaging, Malehmir2021Sparse3Dreflectionseismicsurvey}. This process involves denoising and enhancing the signal quality, which becomes essential due to the nature of the seismic acquisition, as the acquired data is usually corrupted by noise, which increases the complexity of interpretation tasks.

Seismic noise has been studied and classified in the state-of-the-art into random noise (non-signal-dependent) and coherent noise (signal-dependent). Random noise is related to artifacts produced by environmental and equipment-related factors. This category includes noises such as spike-like noise in the seismic marine acquisition, which is characterized by intermittent, impulse-like disturbances. These are typically associated with large marine fauna, such as tuna, that occasionally interfere with the receivers \citep{Volodya2021Noisetypes}; cross-feed noise, which is caused by faulty sensors and resembles noise in the form of stripe-like traces \citep{Volodya2021Noisetypes}. Coherent noise is associated with the corruption of the signal during the data processing stage \citep{mrigya2023convolutional, Yang2023randomnoiseattenuationDnCNN}, it includes smile and frown artifacts produced by over-migration and under-migration of the data, respectively, and are linked to wave propagation phenomena \citep{YOO2022123}; ground roll, which is a common coherent noise in land field seismic data it is a Rayleigh-type surface wave that masks relevant seismic events \citep{jia2024groundrollseparationlandseismic}. An analysis of these types of noise reveals that seismic data is inherently corrupted. As coherent noise is associated with wave propagation, it could be misinterpreted as relevant signals even though coherent noise poses a greater risk to the accuracy of geophysical analysis during interpretation. The literature on denoising predominantly focuses on random noise, as it is more manageable to attenuate due to its consistency with known statistical distributions \citep{Gao2024MultiscaleResidualConvolution,Liu2024SeismicRandomNoiseSuppression,Zhang2024,Wang2024QuadraticUnet}. However, these distributions fail to capture the full complexity of seismic data, which can vary significantly based on site geology, data acquisition methods, and processing techniques. This bias in state-of-the-art denoising solutions highlights an opportunity to address noise outside these known distributions, enabling more accurate representations of seismic noise complexity.

A practical approach to handling high-complexity noises involves building representations that emulate field noise behavior, enabling its subtraction to obtain a clean sample. Image processing techniques create these representations by analytically defining noise as degradation models that corrupt clean data \citep{YOO2022123}. This approach is commonly employed to address noise in poststack seismic data \citep{YOO2022123,WU2022110431,mrigya2023convolutional}.

Several methods have been developed in the literature based on noise emulation to enhance the quality of the poststack seismic data, from filters such as median and band-pass to state-of-the-art deep learning-based solutions. However, classic methods such as frequency-dependent noise attenuation \citep{Al-Heety2022} and structurally oriented coherent noise filtering \citep{dorn2018structurally} depend on known distributions and established models like Gaussian. These approaches assume that seismic noise behaves similarly to these mathematical models, which could work to some extent. However, the complexity of seismic noise often goes beyond these models due to the number of variables involved in acquiring and processing seismic data \citep{Kumar2021Seismic}.

Deep learning-based solutions have shown significant improvements in handling diverse representations of seismic noise for denoising, often relying on supervised learning. This approach trains a denoising network using pairs of clean and noisy patches \citep{mrigya2023convolutional,Hongping2022Denoising}. Nonetheless, these solutions also present a disadvantage: Deep learnig (DL) is data-driven, which implies that the method's performance depends on the amount and distribution of the training data. Typically, a static database is used \citep{Gao2024MultiscaleResidualConvolution,Hongping2022Denoising, mrigya2023convolutional}, where the data remains unchanged during training. On the other hand, unsupervised and self-supervised learning approaches, such as Deep Image Prior (DIP) or internal learning, have been explored to train enhancement networks without relying on external data \citep{Qian2024Unsupervised,Wang2024Self-Supervised}. DIP and internal learning only depend on incomplete measurements to enhance seismic data. However, these methods often struggle with generalization because they require training for each specific data set. In the case of DIP, it is also necessary to know the degradation model or mask that affects the image \citep{Ulyanov_2020}. Therefore, the main limitations are as follows: For classic methods, the dependence on mathematical models that fail to represent the complexity of field noise. For DL solutions, large and variable training data is required, and the static database limits the learning domain, leading to poor generalization across multiple data sets.

This work proposes an approach that improves generalization by adding variability to the database during training. This ensures the method is non-dependent on specific known distributions while avoiding using external data sets, often limited by data ownership in the industry. Our novel approach exploits the advantages of data augmentation to increase the diversity of traditional training databases. 

The proposed method consists of two key processes: The dynamic database, which integrates a Generative Adversarial Network (GAN), and a degradation model to generate clean and corrupted poststack seismic patches, respectively, facilitating data augmentation through the application of the degradation model. The dynamic database guides the supervised enhancement task learning. In addition, to address the possibility of the network encountering unknown noise specific to a geological configuration, we employ domain adaptation via neural style transfer, as described by \cite{neural_style_transfer_2}. This approach allows the model to adapt to new noise patterns beyond the training domain while retaining previously acquired knowledge. 

To summarize, our main contributions are as follows:
\begin{enumerate}[itemsep=0cm]
  \item We propose a supervised preconditioning method for poststack seismic data, leveraging a dynamic guided learning workflow. This method employs a high-variability training strategy based on dynamic data augmentation, enhancing the generalization capabilities of the network to unseen data with diverse noise characteristics.
  \item We introduce a dynamic guided learning scheme with domain adaptation to handle new noise patterns effectively. The method minimizes the mismatch between test and training domains by utilizing a neural style transfer strategy to extract new noise patterns. This approach ensures independence from prior noise distributions and facilitates rapid fine-tuning to adapt to new geological conditions.
  \item We validate the proposed method on synthetic and field seismic data, benchmarking its performance against classic and DL solutions. The results demonstrate that our approach consistently outperforms state-of-the-art methods by effectively capturing complex geological structures and adapting to diverse noise conditions, highlighting its superior robustness and accuracy.
\end{enumerate}

\section{Dynamic Guided Learning}\

\begin{figure}[ht]
    \centering
   \includegraphics[width=1\columnwidth]{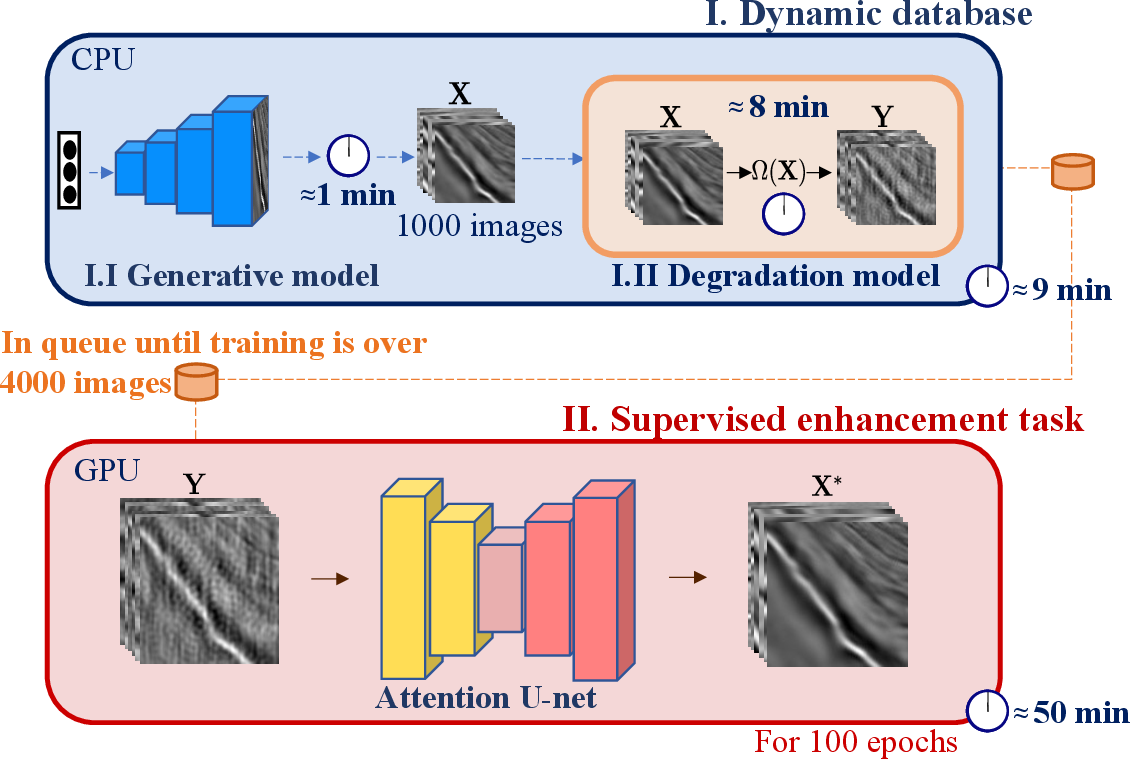}
    
    \caption{The dynamic guided learning workflow with two main processes: (I) the dynamic database that generates $\mathbf{X}$ poststack seismic data using the generative model (1000 patches) and $\mathbf{Y}$ noisy poststack seismic data (4000 patches) using the degradation model containing 12 different types of noise. (II) The supervised enhancement task learns distinctive features of poststack seismic data during training. While process (II) is training, process (I) generates the next batch of images with different types of noise randomly selected, completing one cycle, with (I) in CPU and (II) in GPU to maximize computational efficiency.}
    \label{fig:method}
\end{figure}

Unlike the classic supervised approach, which relies on a static database with a fixed number of images and labels, the dynamic guided learning approach introduces variability. This allows for a more flexible training strategy. Dynamic guided learning consists of two processes, described in the following subsections: the dynamic database (I) and the supervised enhancement task (II) in a workflow shown in Figure \ref{fig:method}. The first process involves generating clean patches using a generative model, denoted as $\mathbf{X}\in \mathbb{R}^{M \times N}$, where $M$ corresponds to the time/depth samples and $N$ represents the trace number. Noisy patches $\mathbf{Y}$ are then created by applying a degradation model to the generated clean patches $\mathbf{X}$. This degradation step acts as a form of data augmentation and constitutes the dynamic database. The second process, the supervised enhancement task, is trained using the dynamically generated data from the database, guiding the learning of the task. Moreover, this workflow operates in cycles, where the dynamic database generates new training data on the CPU while the enhancement network is trained on the GPU. A cycle begins with data synthesis and ends when a predefined number of epochs is completed. These parallel processes ensure maximum efficiency in training by utilizing only the available GPU for model training while data generation occupies the CPU.  

The proposed method focuses on learning the representation of local signal structures and scale-variable seismic noise patterns through a dynamic database, overcoming the limitations of deep learning-based seismic enhancement methods. This improves the generalization capability by feeding the enhancement network with diverse images, as explained in the following subsections.

\vspace{1cm}

\subsection{Dynamic database}\

The first process of the proposed method consists of two components: a generative and a degradation model.

\subsubsection{Generative model}\
To identify the most suitable generative model in terms of computational efficiency and generation quality, we evaluated the Fréchet Inception Distance (FID) metric and generation time on a CPU for the following approaches, as shown in Table \ref{table:gancomp}: Variational Autoencoder GAN (VAE/GAN) \citep{larsen2016autoencoding}, Vector Quantized Variational Autoencoder (VQ-VAE) \cite{van2017neural}, Wasserstein GAN (WGAN) \citep{arjovsky2017wassersteingan} and the Progresive Growing GAN (PGGAN) \citep{karras2018progressivegrowinggans}. It is important to note that we prioritized a balance between computational efficiency and the quality of generated samples. Given that GAN-based models satisfy these criteria, we focus our analysis on this class of models.

\begin{table}[h]
\centering
\caption{Comparison of poststack seismic data generation quality, evaluating different models based on their FID scores and computational efficiency in terms of generation time on CPU. Bold numbers are the best performance while underlined numbers are the second best.}
\label{table:gancomp}

\begin{tabular}{lcccc}
\midrule[2pt]
\bf{Model}    & $\uparrow \bf{IS}$ &  $\downarrow \bf{FID}$ & $\downarrow  \bf{MMD}$ & \bf{Time} (s)                                    \\ \midrule[1pt]
VAE/GAN & 1.148 & 58.244  & 1.131 &$\mathbf{12}$  \\ \hline
VQ-VAE    & 1.000 &  \underline{19.027}    & 0.090 &  $1200$          \\ \hline
WGAN-GP     &    1.004       &  60.738 & $\mathbf{0.054}$   &152  \\ \hline
PGGAN   & $\mathbf{1.201}$ &       $\mathbf{ 10.250}$  &\underline{0.300} &\underline{65}     \\ \midrule[1pt]

\end{tabular}
\end{table}

The selected model was PGGAN as it achieved the best performance in the FID and IS metrics. Although PGGAN did not achieve the best performance in the MMD metric, it remained within a range close to zero. This result suggests that PGGAN exhibits high variability in synthesizing poststack seismic images. Therefore, while the numerical outcome is not the best, it is the second best in the MMD metric, indicating that the generated patches remain well-aligned with the post-stack seismic training distribution.

The generative neural network model $\mathcal{G}$, employs a network that learns the distribution of training data, starting from a low resolution of $K \times L$. It progressively adds layers until the required resolution $M \times N$ is achieved, where $M>
K$ and $N>
L$. The model is conditioned on a fixed normal distribution of latent vectors $\pmb{z} \sim \mathcal{N}(0, \mathbf{I})$ where $\mathbf{I}$ is an identity matrix, enabling the trained model to produce $\mathbf{X}$:

\begin{equation}
\label{eq:gensamp}
    \mathbf{X} = \mathcal{G}(\pmb{z}). 
\end{equation}

\begin{figure}[H]

\centering
\includegraphics[width=\columnwidth]{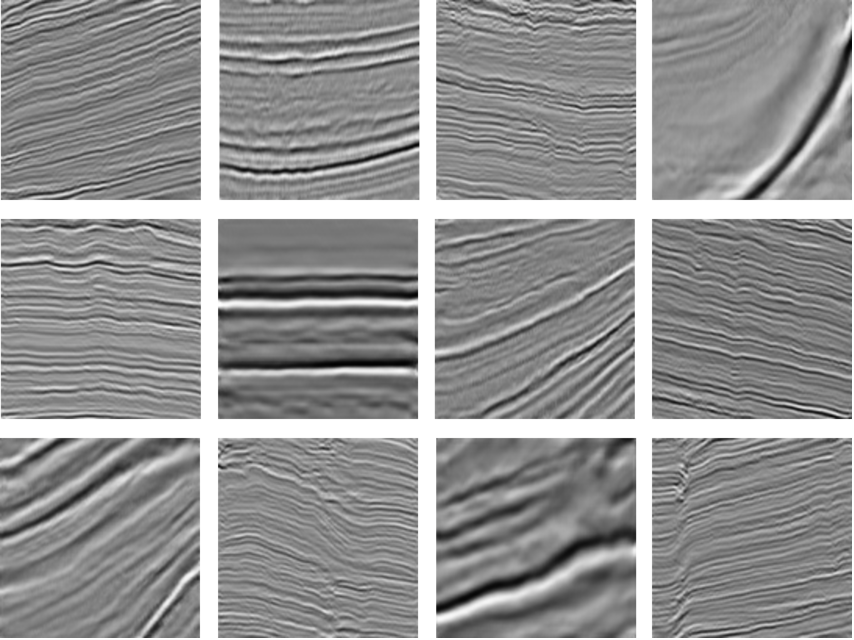}%

\caption{Examples of generated clean $128 \times 128$ poststack seismic patches.}
\label{fig:clsgen}
\end{figure}

Figure \ref{fig:clsgen} shows the generated clean poststack seismic patches coherent with field seismic scenarios based on the expertise of the authors with geology-based knowledge and the evaluated metrics.

\subsubsection{Degradation model}\

\begin{figure*}
\centering
\includegraphics[width=\textwidth]{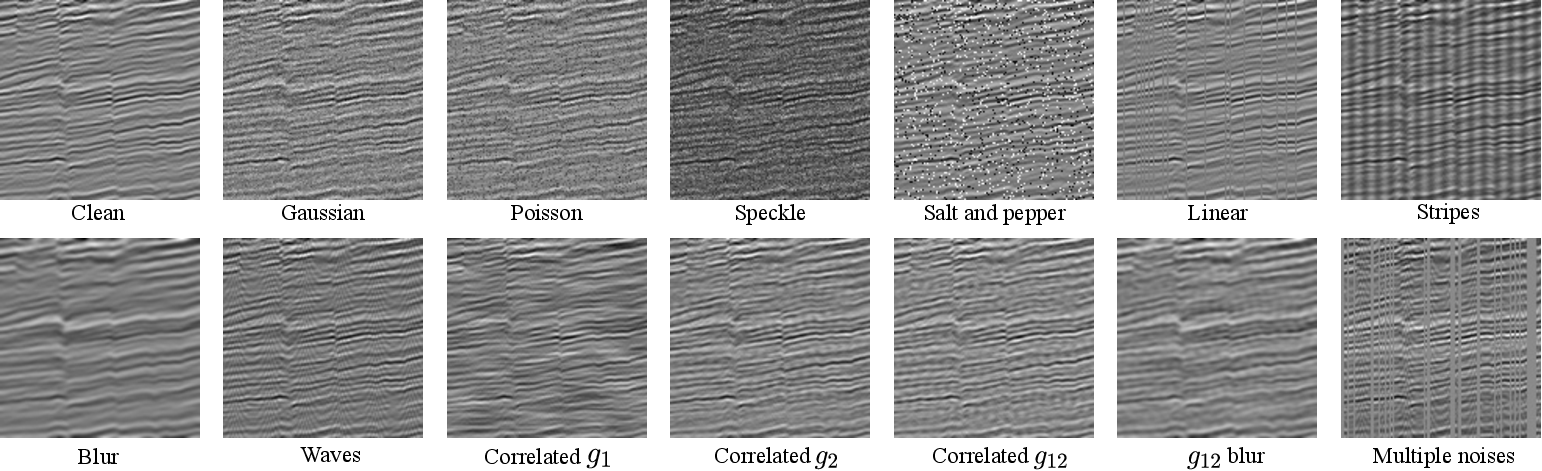}
\caption{Visual example of a clean patch corrupted by the 12 type of noise described in Table \ref{table:tpns}. Note that multiple noise refers to a random combination of three types of noise. }
\label{fig:noise}
\end{figure*}

To generate the inputs $\mathbf{Y}$ of the supervised enhancement task, we employ a degradation model that can be mathematically described as follows:

\begin{equation}
\label{eq:forward}
\mathbf{Y} = \Omega(\mathbf{X}),    
\end{equation}
where $\Omega (\cdot)$ denotes an operator that corrupts the poststack seismic data with $12$ types of noise shown in Table \ref{table:tpns}. Each type of noise was identified by analyzing open-access field poststack seismic data, such as Kerry 3D \citep{kerry}, Blake Ridge Hydrates 3D \citep{blakeridge}, and F3 Netherlands \citep{F3netherlands}. Based on the authors' geophysical expertise, the most common noise types were categorized into 12 main types, with new noise patterns emerging from their combinations. Image processing techniques were employed to emulate these types of noise, as illustrated in Figure \ref{fig:noise}, which shows corrupted poststack seismic data with the characterized noises. 

Each type of noise has a range of weights that varies depending on the implementation. For instance, in the case of Gaussian noise, it is the intensity and goes from $0$, representing no noise, to $1$, indicating the presence of only noise. In the case of linear noise, the parameter corresponds to the number of lines, ranging from $0$, indicating no damage, to the full horizontal extent of the data, where no seismic feature is preserved. These weights allow data augmentation over the generated $\mathbf{X}$ patches, as the model output is a tensor $\mathbf{Y} \in \mathbb{R}^{M \times N \times D}$, where $D$ indicates the number of degradations randomly applied of corrupted patches, each with one randomly selected type of noise, and a combination of multiples noises in the last position. Figure \ref{fig:method} presents an example of this data augmentation, where $1000$ poststack seismic patches are generated, and $D$ is set to 4, increasing the data set to $4000$ noisy patches.
\begin{table}[h]
    \centering
    \caption{Characterization of 12 types of noise related to poststack seismic data, their categorization as coherent or random, and their causes.}
    \begin{tabular}{lll}
        \midrule[2pt]
        \textbf{Noise} & \textbf{Category} & \textbf{Cause} \\ 
        \midrule[1pt]
        Gaussian & Random & Human related \\ 
        \hline
        Poison & Random & Human related \\ 
        \hline
        Speckle & Random & Human related \\ 
        \hline
        Salt and pepper & Random & Human related \\ 
        \hline
        Linear & Random & Cross-feed \\ 
        \hline
        Stripes & Random & Cross-feed \\ 
        \hline
        Blur & Random & Processing \\ 
        \hline
        Waves & Coherent & Smile and artifacts \\ 
        \hline
        Correlated $g_1$  & Coherent & Footprint \\ 
        \hline
        Correlated $g_2$  & Random & Spike-like \\ 
        \hline
        Correlated $g_{12}$  & Coherent & Footprint \\ 
        \hline
        $g_{12}$ blur & Random & Background \\ 
        \midrule[1pt]
    \end{tabular}
    
    \label{table:tpns}
\end{table}

 The dynamic database is used to guide the learning of the supervised enhancement task.
\begin{figure*}[ht]
    \centering
    \includegraphics[width=1\textwidth]{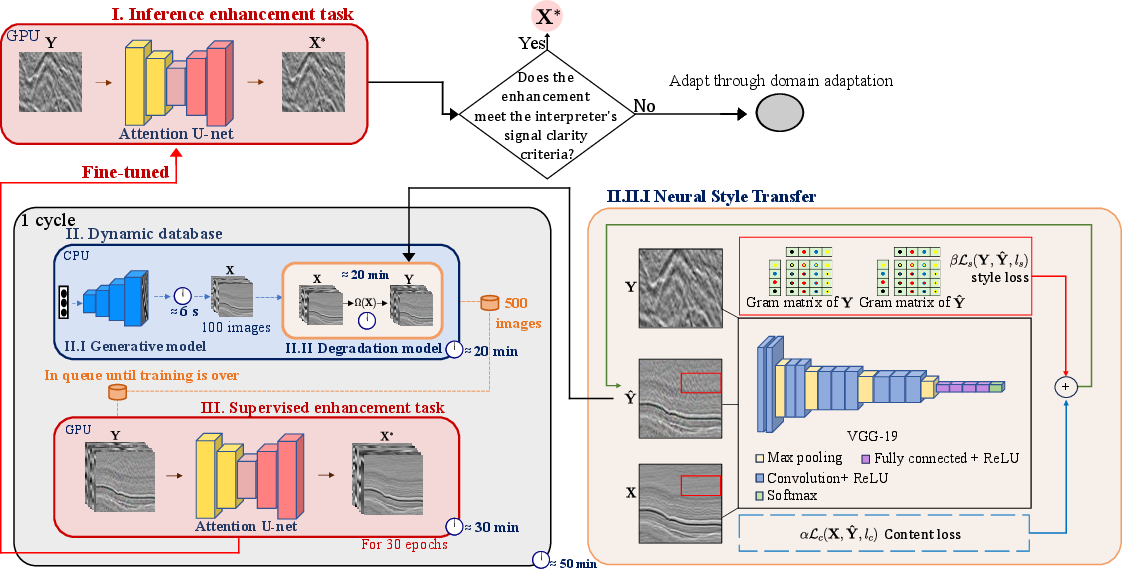}
    \caption{Domain adaptation via neural style transfer. When the inference enhancement task (I) testing does not meet enhancement criteria, the proposed method is employed to fine-tune the Attention U-net and adapt to the unknown noisy domain. The neural style transfer (number II.II.I) is used to adapt the domain and acts as a complement of the degradation model (number II.II), corrupting the $\mathbf{X}$ clean poststack seismic patches with the aim for the network to learn the new domain.}
    \label{fig:domadp}
\end{figure*}
\subsection{Supervised enhancement task}\

The supervised enhancement task employs a supervised learning approach with pairs of $\bf{X}$ and $\bf{Y}$ to train an enhancement network $\mathcal{M_\theta}$ to optimize the parameters $\theta$ by minimizing the loss function:
\begin{equation}
\label{eq:gloss}
\begin{split}
  \theta^* = \argminB_{\theta} \mathcal{L}{(\theta)}, 
\end{split}
\end{equation}
where $\mathcal{L}$ is the cost function described in Equation \ref{eq:loss}.

The $\bf{X}$ and $\bf{Y}$ are generated dynamically. This process is performed recurrently, with the dynamic database preparing a new training batch for every defined number of epochs. This exposes $\mathcal{M_\theta}$ to different sets of noisy patches. The key advantage of this approach is that it increases the variability and adaptability of the training process. 

\subsubsection{Enhancement loss function}\ 

The loss function employed in this work is inspired by methodologies from related studies on seismic data enhancement \citep{goyespenafiel2024gansupervisedseismicdatareconstruction}. It measures the similarity between the original clean seismic data $\mathbf{X}$, as defined in Equation \ref{eq:gensamp}, and the enhanced data $\mathbf{X}^* = \mathcal{M}_\theta (\mathbf{Y})$. Here, 
$\mathbf{Y}$ represents the forward-transformed version of $\mathbf{X}$, obtained using the degradation model $\Omega (\cdot)$, as described in Equation \ref{eq:forward}. To capture this similarity, the loss function combines the Mean Absolute Error (MAE) and the Structural Similarity Index Measure (SSIM), striking a balance between pixel-level fidelity and perceptual image quality. The loss function is defined as:

\begin{equation}
\label{eq:loss}
\begin{split}
    \mathcal{L}(\theta) = \frac{1}{B}\left \| \mathbf{X} - \mathbf{X^*} \right \|_1 + 
    (1 - \mathrm{SSIM}(\mathbf{X},\mathbf{X^*})),
\end{split}
\end{equation}
where the first term, $\parallel \cdot \parallel_1 $, represents the $\ell_1$-norm, which, when divided by $B$ the batch size, results in the MAE, which evaluates the average absolute difference between the clean seismic data $\mathbf{X}$ and the enhanced data $\mathbf{X}^*$. The second term involves the SSIM, which quantifies luminance, contrast, and structural similarity between $\mathbf{X}$ and $\mathbf{X}^*$. It has a maximum value of 1, encouraging the enhancement to maintain perceptual consistency \citep{BAKUROV2022SSIM}.

\subsubsection{Enhancement network configuration}\

The supervised enhancement task uses a 2D U-net with attention blocks \citep{oktay2018attentionunetlearninglook}. Downsampling and upsampling operations are performed using layers with $2 \times 2$  strides, respectively. Skip connections are included between the encoder and the decoder to facilitate the fusion of low-level and high-level features. Before the attention blocks, the input and output of the Attention U-net are concatenated to enhance the spatial information. Batch normalization and the ReLU activation function ensure consistency across all layers. The four attention blocks in the image enhancement process gradually improve the patch quality before reaching a convolutional layer with a single filter and a sigmoid activation function, which generates the enhanced seismic patch. The total number of trainable parameters is $34,877,421$. Further details of the implementation can be found in the project's repository.

\subsection{Domain adaptation via neural style transfer}\

\begin{figure}[ht]
    \centering
    \includegraphics[width=\linewidth]{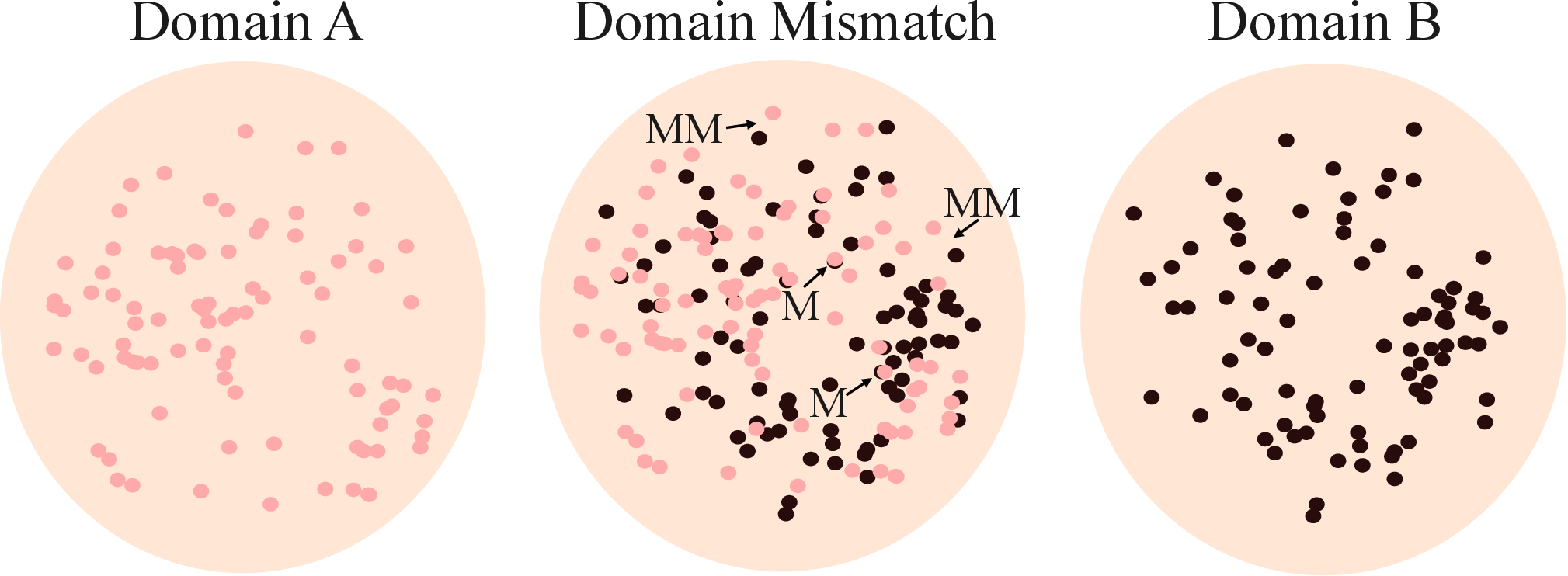}
    \caption{Example of domain mismatch across Domain A, which represents the training poststack seismic data, and Domain B, which represents the test field poststack seismic data. M stands for match, and MM for mismatch. }
    \label{fig:dmnms}
\end{figure}

\begin{figure}[ht]
    \centering
    \includegraphics[width=\columnwidth]{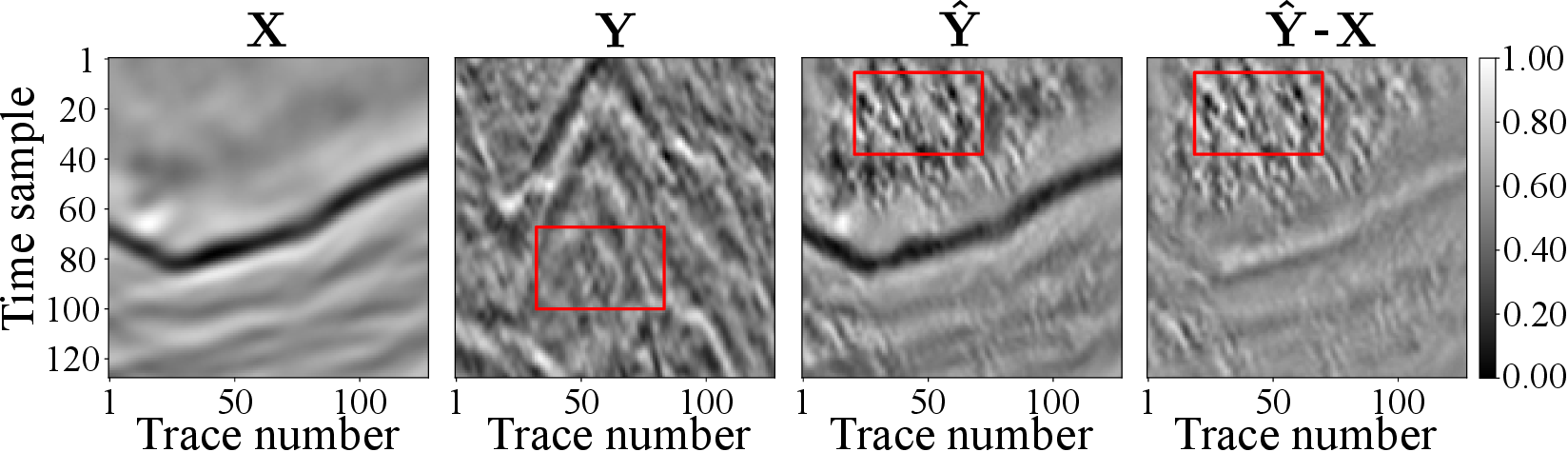}
    \caption{Domain adaptation on a clean poststack seismic patch $\mathbf{X}$ derived from a field noisy patch of the Kerry 3D data set $\mathbf{Y}$. The generated data $\mathbf{\hat{Y}}$ with the content data's structure and the style data's noise. The style difference between $\mathbf{\hat{Y}}$ and $\mathbf{X}$. The red box highlights a noisy area of the field-style data and the noise transfer over the generated patch.}
    \label{fig:transfer}
\end{figure}

The performance of deep supervised learning models during testing is often lower than during the training phase due to biases introduced by the training data domain. This leads to a mismatch between the training and testing domains, as illustrated in Figure \ref{fig:dmnms}. Therefore, the network $\mathcal{M_\theta}$ must be adapted to the new domain. Training from scratch can effectively learn domain-specific features; however, this approach is highly inefficient, especially when dealing with multiple seismic data sets with distinct conditions, such as geological structures, acquisition instruments, or processing methods. Instead, we use a fine-tuning strategy that leverages pre-trained knowledge to adapt the network and learn new seismic features efficiently. To this end, we use an approach for generating field noise data using neural style transfer \citep{neural_style_transfer}. As Figure \ref{fig:domadp} shows, if the enhancement after inference (I) is unsatisfactory, this approach can incorporate the new noise pattern into the degradation model (II.II) as a complement of the 12 characterized types of noise to fine-tune the network following the same workflow as training. The new noise pattern is introduced by extracting the style from the field noisy poststack seismic data $\mathbf{Y}$ and adapting it to the content data clean patches  $\mathbf{X}$ by synthesizing new data $\mathbf{\hat{Y}}$, initialized as $\mathbf{X}$, under the assumption that style features are noise features, using a pre-trained VGG-19 model. This improves generalization and helps resolve the domain mismatch problem. Figure \ref{fig:transfer} shows an example of the style transfer between a clean poststack seismic patch $\mathbf{X}$ and field data $\mathbf{Y}$. The cost function for the optimization problem is expressed as follows:

\begin{equation}
\label{eq:neural}    
\mathcal{L}_{total}= \alpha \mathcal{L}_c (\mathbf{X, \hat{Y}},l_c)+ \beta \mathcal{L}_s(\mathbf{Y}, \mathbf{\hat{Y}},l_s),
\end{equation}
where $\mathcal{L}_c$ is the content loss, calculated over the layer $l_c$, to preserve seismic information between $\mathbf{X}$ and $\mathbf{\hat{Y}}$. Similarly. $\mathcal{L}_s$ is the style loss calculated over the layer $l_s$, to extract the noise from the field image. Noise features are assumed to correspond to style features, derived from the calculation of Gram matrices for both the style and content data; $\alpha$ and $\beta$ are the weights that determine the importance of structure $\mathcal{L}_c$ and style $\mathcal{L}_s$, relative to each other.  

\section{Experiments and results}\

\subsubsection{Data sets}\

The following synthetic and field open-access data sets were used for training: SEAM Phase 1 \citep{seamPhaseI}, TGS salt identification challenge \citep{tgs}, 1994 BP migration from topography \citep{1994bp}, AGL Elastic Marmousi \citep{AGL}, Kerry 3D \citep{kerry}, and F3 Netherlands \citep{F3netherlands}. Patches with dimensions $M = 128$ and $N=128$ were extracted from each data set, ensuring the selection of noise-free patches. Table \ref{table:datasets} shows the number of patches extracted from each data set.

\begin{table}
\centering
\caption{Number of field and synthetic patches for each data set.}
\begin{tabular}{ m{0.3\columnwidth}  c c }
\midrule[2pt]
\textbf{Data set} & \textbf{Type} & \textbf{Number of patches}  \\
\midrule[1pt]

SEAM Phase 1 & Synthetic & 1119  \\ \hline
TGS & Synthetic & 630 \\ \hline
1994 BP & Synthetic & 803  \\ \hline
AGL Elastic Marmousi & Synthetic & 152  \\ \hline
Kerry 3D & Field & 3127 \\ \hline
F3 Netherlands & Field & 2134  \\ \hline
 \textbf{Total} &  & $\mathbf{8000}$  \\ \midrule[1pt]
\end{tabular}

\label{table:datasets}
\end{table}

\subsubsection{Metrics}\

To evaluate the performance of the PGGAN for data generation, state-of-the-art metrics, such as the inception score (IS), Fréchet inception distance (FID), and Maximum Mean Discrepancy Metric (MMD) metrics were used \citep{dash2021reviewgenerativeadversarialnetworks}. The Peak Signal-to-Noise Ratio (PSNR) and SSIM \citep{BAKUROV2022SSIM} metrics were employed to evaluate the enhancement over noisy poststack seismic data. 

\subsubsection{Experimental setup}\ 

The PGGAN model was trained for 770 epochs, following the implementation by \cite{progan}, using the synthetic and field data sets presented in Table \ref{table:datasets}. The enhancement network, $\mathcal{M}_\theta$, was trained for $15$ cycles, each consisting of $100$ epochs. During training, in the dynamic database, the generative model generated $1000$ $\mathbf{X}$ post-stack data. The degradation model was used to augment the data with a factor of $D=4$, resulting in a total of 4000 pairs of clean and corrupted patches, which were processed by $\mathcal{M}_\theta$ in each training cycle. By the end of training, 
$\mathcal{M}_\theta$ had learned from $60,000$ unique patches. This total was derived from the product of the number of cycles ($15$) and the number of patches per cycle ($4000$). The total number of epochs across all training cycles was $1500$. During testing, we found that this configuration of generated patches and the chosen degradation factor 
$D$ yielded optimal performance. While this approach can be extended to a greater $D$ factor or a larger number of generated patches, its selection was constrained by computational limitations. The Adam optimizer \citep{kingma2017adammethodstochasticoptimization} was employed, with a batch size of $B = 200$ and a learning rate set to $10^{-2}$, subjected to an exponential decay of 0.6 every 200 epochs. The neural models were implemented using the PyTorch library. The \textbf{training equipment} was a system with an AMD Ryzen 7 5800X3D CPU for managing the dynamic database and an NVIDIA GeForce RTX 4090 GPU for the supervised enhancement task. Under these settings, the training of the $15$ cycles took $15$ hours to complete, required $9$ minutes to generate the $4000$ patches, and $50$ minutes for the $100$ epochs of training per cycle. The style transfer loss weights for domain adaptation were set to $\alpha = 1$ and $\beta = 50$. The \textbf{testing equipment} consisted of an Intel(R) Core(TM) i7-10700K CPU and an NVIDIA GeForce RTX 3080 ti GPU to demonstrate computational efficiency.

To demonstrate the effectiveness of the proposed method, we conducted four experiments in local and complete poststack seismic sections.  

\subsection{Experiment I}\

The enhancement network, trained using the dynamic guided learning workflow, was evaluated on 2D local poststack seismic data with spatial dimensions of $128 \times 128$. In this experiment, a clean seismic patch was corrupted using the $12$ characterized types of noise to prove the enhancement efficiency over the known noise domain.
 \begin{figure}[ht]

    \centering
    \includegraphics[width=\columnwidth]{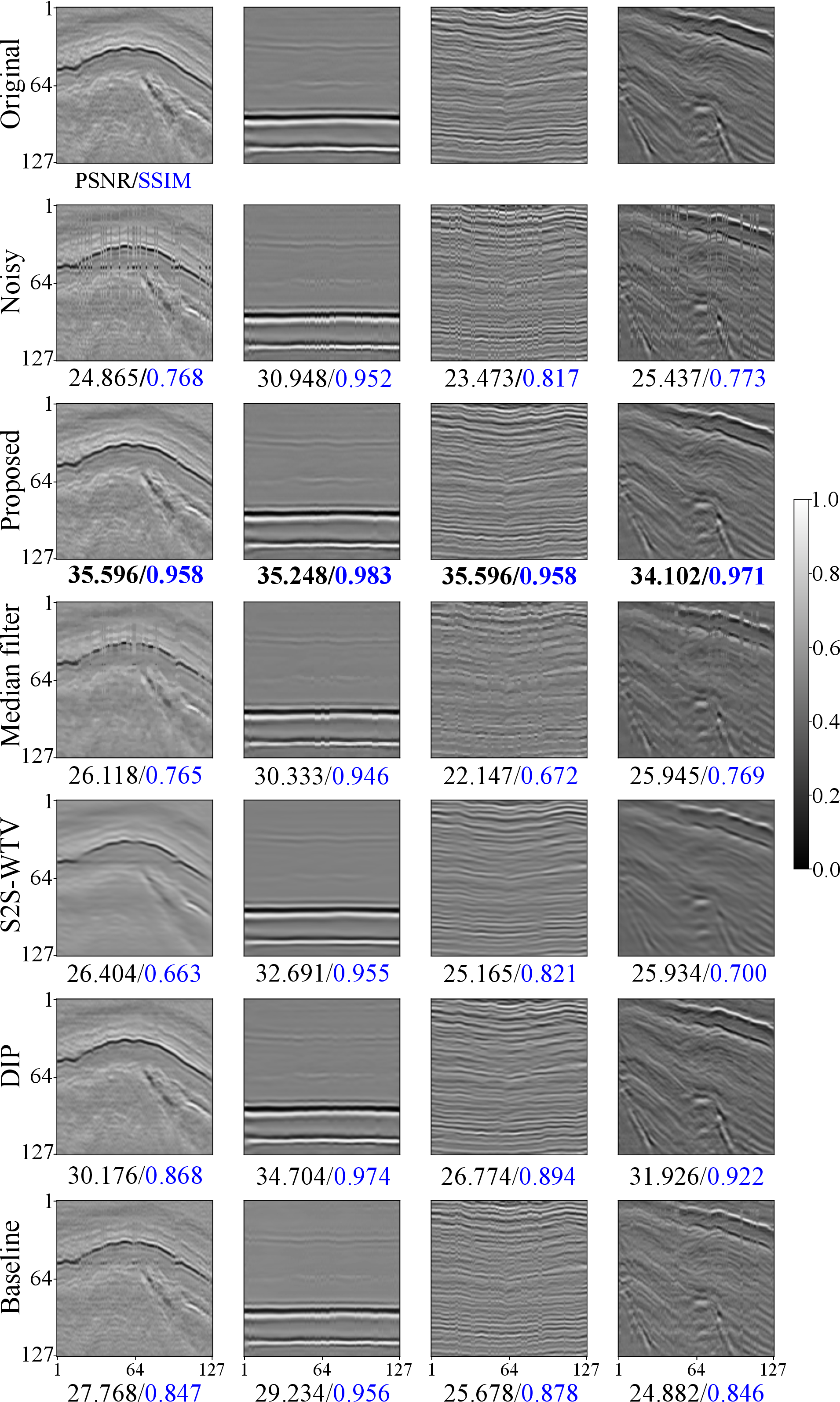}
    \caption{Clean-generated poststack seismic patches (original) corrupted (noisy) by linear noise, along with a comparison of the enhancement achieved using the proposed method and the state-of-the-art solutions, using PSNR (dB) and SSIM metrics. Bold numbers indicate the best result of the enhancement.}
    \label{fig:expI}
\end{figure}
\begin{table*}

\centering
\caption{Comparison metrics from experiment I. PSNR (dB) and SSIM metrics across different simulated types of noise for enhancement methods, including the median filter, DIP, S2S-WTV, and Baseline. Bold text indicates the best result, while underlined text highlights the second-best.}
\begin{tabular}{ccccccccccc}
\midrule[2pt]
Noises               & \multicolumn{2}{c}{Medial filter}                   & \multicolumn{2}{c}{DIP}                             & \multicolumn{2}{c}{S2S-WTV}                         & \multicolumn{2}{c}{Baseline}                        & \multicolumn{2}{c}{Proposed}                        \\ \cline{2-11} 
\multicolumn{1}{l}{} & \multicolumn{1}{l}{SSIM} & \multicolumn{1}{l}{PSNR} & \multicolumn{1}{l}{SSIM} & \multicolumn{1}{l}{PSNR} & \multicolumn{1}{l}{SSIM} & \multicolumn{1}{l}{PSNR} & \multicolumn{1}{l}{SSIM} & \multicolumn{1}{l}{PSNR} & \multicolumn{1}{l}{SSIM} & \multicolumn{1}{l}{PSNR} \\ \midrule[1pt]
Gaussian             & 0.783                    & 22.947                   & \underline{0.965}                    & \underline{31.992}                   & 0.952                    & 29.021                   & 0.975                    & 31.658                   & \textbf{0.985}           & \textbf{33.457}          \\ \hline
Poisson              & 0.784                    & 22.899                   & \underline{0.961}                    & \underline{31.997}                   & 0.952                    & 28.493                   & 0.974                    & 31.291                   & \textbf{0.986}           & \textbf{34.248}          \\ \hline
Speckle              & 0.563                    & 15.787                   & \underline{0.908}                    & \underline{27.273}                   & 0.781                    & 16.804                   & 0.807                    & 19.370                   & \textbf{0.934}           & \textbf{27.336}          \\ \hline
Salt and pepper      & 0.712                    & 20.833                   & \underline{0.881}                    & \underline{28.028}                   & 0.731                    & 20.742                   & 0.825                    & 22.551                   & \textbf{0.992}           & \textbf{35.922}          \\ \hline
Linear               & 0.560                    & 18.867                   & \underline{0.947}                    & \underline{29.238}                   & 0.839                    & 23.620                   & 0.926                    & 26.580                   & \textbf{0.994}           & \textbf{36.859}          \\ \hline
Waves                & 0.770                    & 22.124                   & 0.957                    & 30.986                   & 0.949                    & 27.884                   & \underline{0.976}                    & \underline{31.602}                   & \textbf{0.991}           & \textbf{35.319}          \\ \hline
Stripes              & 0.633                    & 18.838                   & \underline{0.964}                    & \underline{32.717}                   & 0.798                    & 20.687                   & 0.948                    & 25.201                   & \textbf{0.998}           & \textbf{39.361}          \\ \hline
Correlated $g_1$     & 0.782                    & 22.774                   & 0.944                  & \underline{29.538}                   & 0.934                    & 27.272                   & \underline{0.965}                   & 29.442                   & \textbf{0.972}           & \textbf{31.108}          \\ \hline
Correlated $g_2$     & 0.784                    & 22.923                   & 0.950                    & \underline{31.654}                   & 0.942                    & 28.509                   & \underline{0.978}                    & 31.001                   & \textbf{0.987}           & \textbf{34.262}          \\ \hline
Blur                 & 0.516                    & 17.400                   & \textbf{0.904}           & \textbf{30.426}          & 0.550                    & 17.707                   & 0.600                    & 19.085                   & \underline{0.901}                    & \underline{27.337}                   \\ \hline
Correlated $g_{12}$  & 0.796                    & 23.124                   & \underline{0.954}                    & \underline{31.062}                   & 0.953                    & 28.775                   & 0.981                    & 31.271                   & \textbf{0.984}           & \textbf{33.593}          \\ \hline
$g_{12}$ blur        & 0.683                    & 19.403                   & \underline{0.948}                   & \underline{30.456}                   & 0.770                    & 20.496                   & 0.851                    & 22.129                   & \textbf{0.981}           & \textbf{31.543}          \\ \midrule[1pt]
\end{tabular}
\label{table:psnr-ssim}
\end{table*}
\begin{figure*}
{\includegraphics[width=\textwidth]{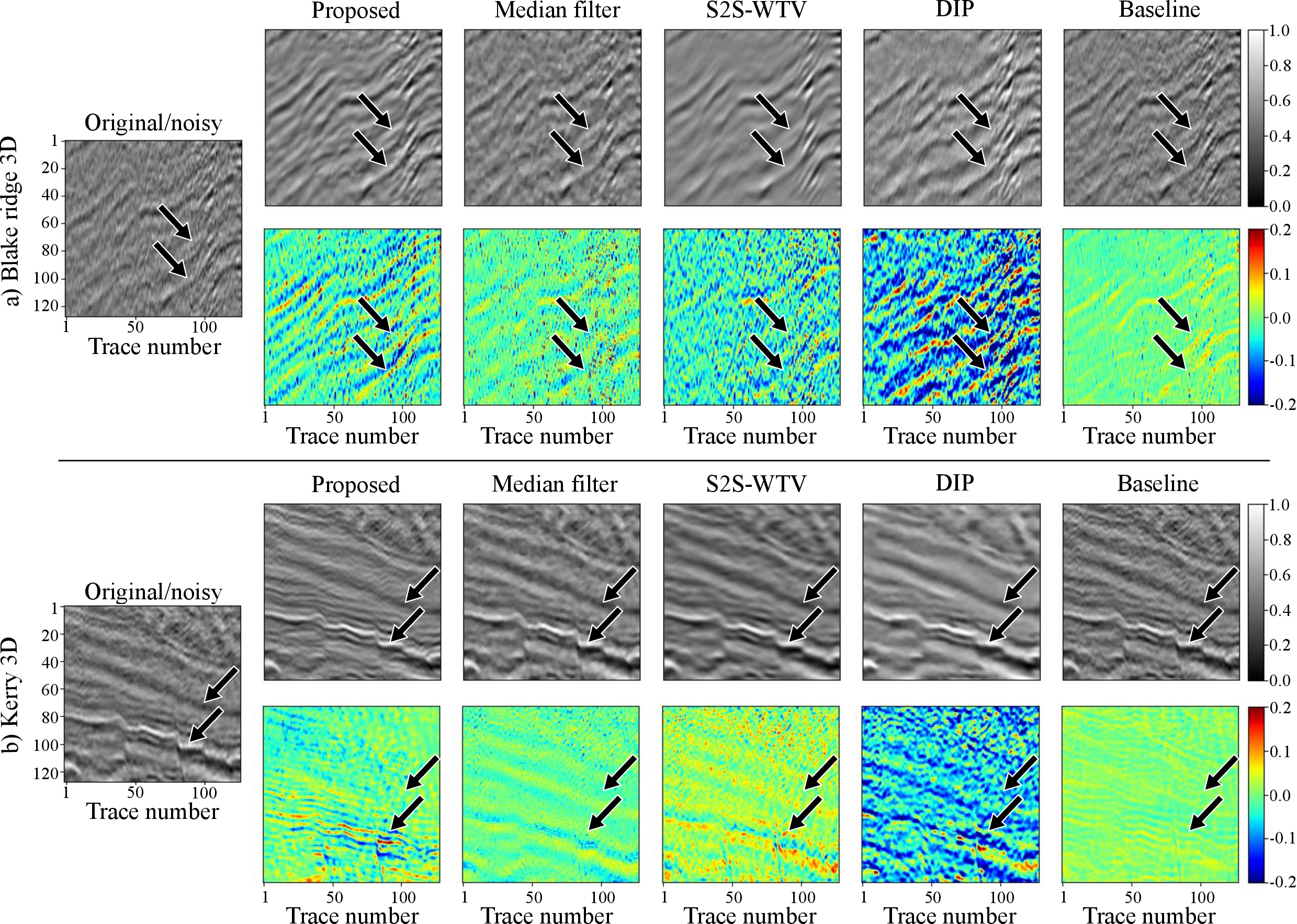}}%
\caption{Visual results of experiment II. Local $128 \times 128$ patches from a) Blake Ridge 3D and b) Kerry 3D data sets enhanced compared against state-of-the-art methods and the difference between the original and enhanced data.}
\label{fig:br128s}
\end{figure*}
\begin{figure*}[ht]

{\includegraphics[width=\textwidth]{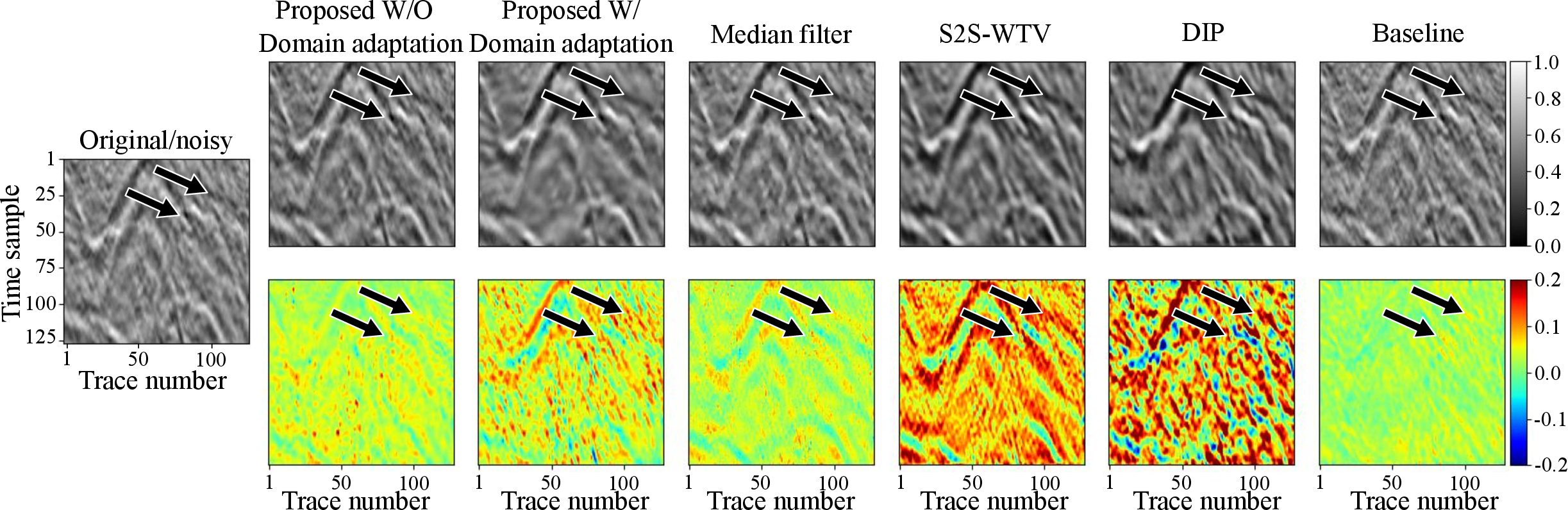}}%

\caption{Visual results of experiment III. Local patches from the Kerry 3D data set enhanced with (W/) and without (W/O)  domain adaptation compared against state-of-the-art methods and the difference between the original and enhanced data.}
\label{fig:fine128s}
\end{figure*}
Three distinct noise weights, low, medium, and high, were assigned for each noise type. Each noise was applied $500$ times for every weight, resulting in $1500$ different instances of corrupted data per type of noise due to the stochastic nature of the noise generation process. The method's performance was evaluated by calculating the average PSNR and SSIM metrics across the $1500$ corrupted instances for each type of noise, taking an average enhancement time of $8$ milliseconds per patch. We also compared the method with other state-of-the-art enhancement methods, such as the median filter, which attenuates noise by replacing each pixel value with the median of its neighboring pixels \citep{George2018Surveymedian}; self2self weighted total variation (S2S-WTV), which employs an optimization-based method to attenuate noisy signals \citep{Xu2023s2s-wtv}; DIP \citep{Ulyanov_2020}; and the baseline, which refers to the enhancement network of the proposed method trained using a classical supervised learning approach for $1500$ epochs. The training process utilized a static database comprising clean and noisy patches, where the noisy patches were generated from the 12 characterized types of noise. 

Figure \ref{fig:expI} illustrates the process of corrupting a patch and the enhancement using the proposed method, demonstrating its advantage in preserving the signal integrity of seismic reflection interfaces, something that other methods, such as DIP and S2S-WTV, fail to achieve due to over-smoothing. In contrast, the baseline and median filter methods struggle with effective noise attenuation. Table \ref{table:psnr-ssim} summarizes the results for all types of noise. This experiment shows that our method achieves an average improvement over the state-of-the-art solutions of up to $8$ dB in PSNR and $0.127$ in SSIM for noise attenuation across the characterized types of noise.

It is worth noting that even though the proposed method exhibited superior performance, it was outperformed only by DIP under Blur-type noise by 3 dB in PSNR, as shown in Table \ref{table:psnr-ssim}. This discrepancy may be attributed to DIP's reliance on a noise mask, which assumes prior knowledge of noise characteristics. However, this represents an ideal scenario not met in real-world applications, as the noise in field poststack data is often unknown. Nonetheless, the proposed approach still achieved the second-best performance compared to the ideal case.

\subsection{Experiment II}\

This experiment used $128 \times 128$ patches of 2D poststack seismic field data for testing. The ground truth for this data is unavailable, highlighting the method's application in a field scenario. Figure \ref{fig:br128s} illustrates the experimental results on patches from the Kerry 3D and Blake Ridge 3D data sets. 

As shown in Figure \ref{fig:br128s}, the patch from Blake Ridge 3D exhibits linear noise that affects the continuity of seismic events, while the patch from Kerry 3D contains a random-like noise resembling a Gaussian distribution. After training with the dynamic guided learning workflow, the enhancement network effectively improves the signal in local patches while preserving the coherence of seismic structures, outperforming comparable approaches such as DIP and S2S-WTV. While these methods occasionally yield good results, they often lead to over-smoothing of the seismic signal, as seen in the Blake Ridge patch. In contrast, the proposed approach avoids this issue, preserving the seismic reflection events highlighted by the black arrows in the Kerry and Blake Ridge patches. Notably, DIP struggles due to its reliance on a noise mask, which is typically unavailable in real-world applications. This underscores the proposed method's ability to maintain structural integrity by distinguishing and highlighting distinct features of poststack seismic data, making it more reliable across diverse data sets. 
\begin{figure*}[t]
    \centering
    \includegraphics[width=\linewidth]{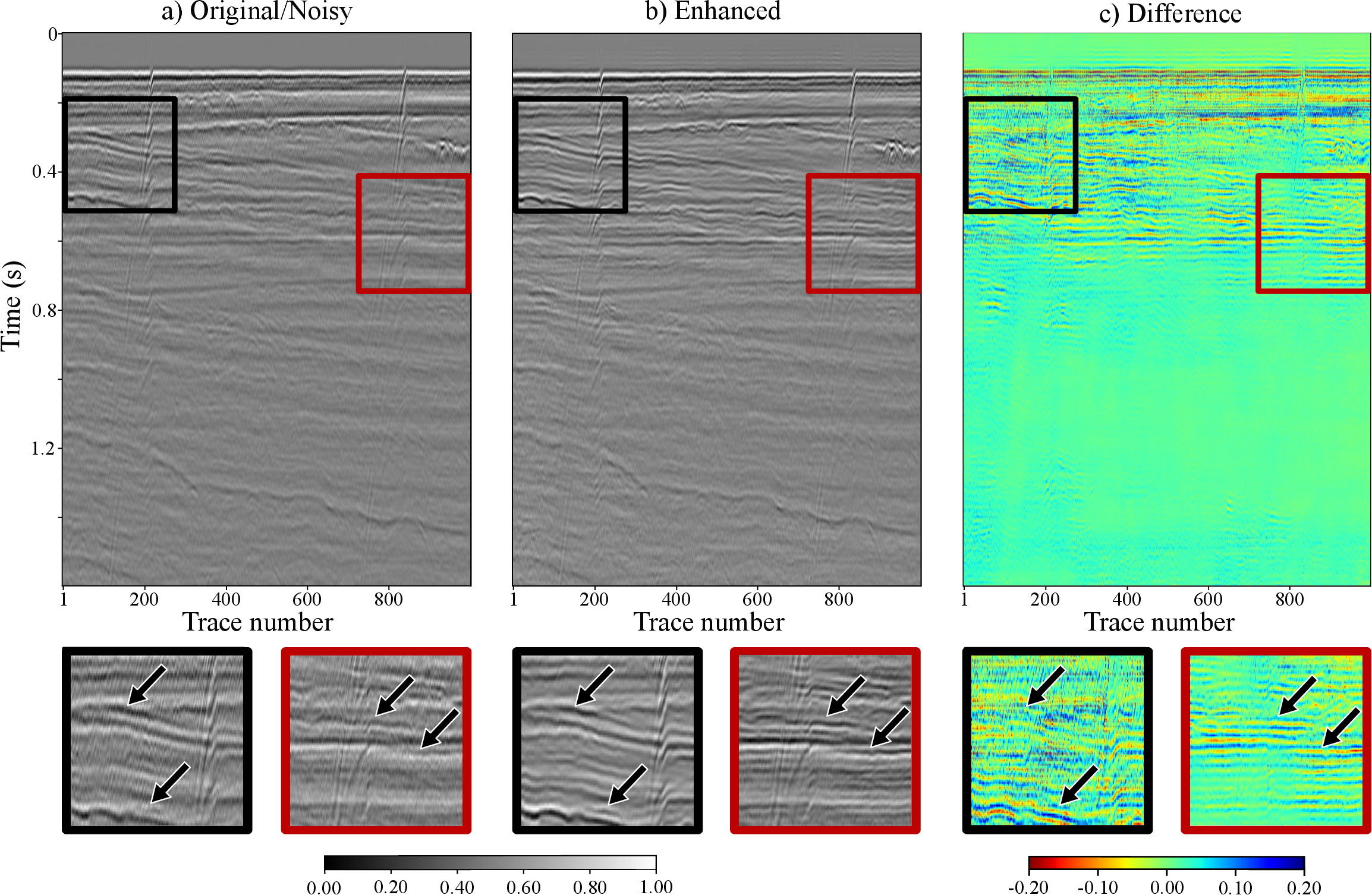}
    \caption{Visual results of experiment IV using the Mobil AVO Viking Graben data set. a) the original data with noise artifacts, (b) the enhanced complete section and c) the difference between a) and b). Zoomed-in views of regions black and red highlight the main enhancement features.}
    \label{fig:expiv}
\end{figure*}
\subsection{Experiment III}\

Although training with the dynamic guided learning workflow effectively enhances seismic signals, there remains a possibility of encountering seismic data outside the training domain. The 12 characterized types of noise provide a robust representation of potential seismic noise scenarios. However, due to complex geological settings, they may not cover the full spectrum of possible seismic noise conditions. To address this issue, we employ domain adaptation using neural style transfer to learn new noise patterns. 

In this experiment, the enhancement of a patch from the Kerry 3D data set containing a noise type unknown to the network was unsatisfactory. Consequently, we applied the domain adaptation strategy. Figure \ref{fig:fine128s} illustrates the results with (W/) and without domain adaptation (W/O). The W/ results were achieved by fine-tuning the network for 30 epochs. The black arrows highlight that while the proposed method successfully attenuates noise by focusing on the diagonal noise pattern after domain adaptation, S2S-WTV and DIP over-corrected some signals. In contrast, the median filter and baseline methods failed to reduce the noise level effectively.

\subsection{Experiment IV}\

The results presented in the previous sections demonstrate that dynamic guided learning is effective for training the supervised enhancement task and outperforms state-of-the-art solutions. To evaluate the method's applicability in field poststack seismic sections used in seismic processing workflows, we conducted tests over the Mobil AVO Viking Graben data set \citep{mobil_avo_1994}. It is worth noting that this data set was not used during training, emphasizing the robustness of the method's generalization ability. 

The enhancement network was trained using patches with spatial dimensions of $128 \times 128$. To enhance poststack seismic sections exceeding the training dimensions, the sections were divided into patches that matched the training size, incorporating overlapping regions. Each patch was enhanced individually, and the enhanced data was reconstructed by taking the median of the overlapping regions. This approach was designed to preserve as many relevant signals as possible. The entire process, including patch division, enhancement, and reassembly of the seismic section for the Mobil AVO Viking Graben data set, took 18 seconds. This demonstrates that the method effectively enhances poststack seismic data and is computationally efficient. Figure \ref{fig:expiv} showcases the results, visually comparing the original noisy data and the enhanced version.

Figure \ref{fig:expiv}a shows common artifacts in poststack seismic data, such as random high-frequency noise. The red box highlights a region with this particular noise pattern that masks structural details in the original data. Such noise degrades the visual quality of seismic images, affecting their analysis. Our method, illustrated in Figure \ref{fig:expiv}b, effectively attenuates this noise, leading to more evident seismic reflections. The enhanced data further demonstrates improved continuity and better feature visibility over the black box. Notably, regions emphasized by black arrows across all boxes illustrate areas where significant improvements have been achieved, enhancing the overall coherence of the complete section.

\section{Conclusions}\

This work presents a robust and adaptable method for data preconditioning in seismic analysis. Given the significant variability of seismic noise across different geological settings, our approach provides a reliable solution for mitigating noise-related challenges, thereby enhancing the accuracy and reliability of seismic interpretation in industrial applications.

Our method demonstrates independence from specific seismic noise domains. Although the network was initially trained on 12 characterized types of noise, it could adapt during testing, effectively addressing field-specific and previously unknown noise domains that challenge the enhancement task. This adaptability underscores the method's applicability to synthetic and field data sets outside the training domain.

\begin{acknowledgments}\
This work was funded by the Vicerrectoría de Investigacion y Extensión from Universidad Industrial de Santander under Project 3925. The authors acknowledge the High Dimensional Signal Processing Group (HDSP) from Universidad Industrial de Santander for providing the computational resources to run the experiments.
\end{acknowledgments}

\section*{Data and Materials Availability}
For reproducibility, the data and codes to replicate the proposed method are available at \url{https://github.com/TJaqui/Seismic_denoising_scheme}

\bibliographystyle{seg}  
\bibliography{example}

\begin{thebibliography}{}
\itemsep0pt

\bibitem[Al-Heety and Thabit, 2022]{Al-Heety2022}
Al-Heety, A. J.~R., and H.~A. Thabit,  2022, {Random and coherent noise attenuation for 2D land seismic reflection line acquired in Iraq}: Geophysical Prospecting,  337--354.

\bibitem[Arjovsky et~al., 2017]{arjovsky2017wassersteingan}
Arjovsky, M., S. Chintala, and L. Bottou,  2017, {Wasserstein generative adversarial networks}: International conference on machine learning, PMLR, 214--223.

\bibitem[Bakurov et~al., 2022]{BAKUROV2022SSIM}
Bakurov, I., M. Buzzelli, R. Schettini, M. Castelli, and L. Vanneschi,  2022, {Structural similarity index (SSIM) revisited: A data-driven approach}: Expert Systems with Applications, {\bfseries 189}, 116087.

\bibitem[Baroni et~al., 2018]{F3netherlands}
Baroni, L., R.~M. Silva, R. S.~Ferreira, D. Chevitarese, D. Szwarcman, and E. Vital~Brazil,  2018, Netherlands {F}3 interpretation dataset.

\bibitem[Chopra and Marfurt, 2013]{Chopra2013}
Chopra, S., and K.~J. Marfurt,  2013, {Preconditioning seismic data with 5D interpolation for computing geometric attributes}: The Leading Edge, {\bfseries 32}, 1456--1460.

\bibitem[Dash et~al., 2024]{dash2021reviewgenerativeadversarialnetworks}
Dash, A., J. Ye, and G. Wang,  2024, {A Review of Generative Adversarial Networks (GANs) and Its Applications in a Wide Variety of Disciplines: From Medical to Remote Sensing}: IEEE Access, {\bfseries 12}, 18330--18357.

\bibitem[Dorn, 2018]{dorn2018structurally}
Dorn, G.~A.,  2018, {Structurally Oriented Coherent Noise Filtering}: First Break, {\bfseries 36}, 37--45.

\bibitem[Du et~al., 2022]{neural_style_transfer_2}
Du, H., Y. An, Q. Ye, J. Guo, L. Liu, D. Zhu, C. Childs, J. Walsh, and R. Dong,  2022, {Disentangling Noise Patterns From Seismic Images: Noise Reduction and Style Transfer}: IEEE Transactions on Geoscience and Remote Sensing, {\bfseries 60}, 1--14.

\bibitem[Fehler and Keliher, 2011]{seamPhaseI}
Fehler, M., and P.~J. Keliher,  2011, {SEAM Phase 1: Challenges of Subsalt Imaging in Tertiary Basins, with Emphasis on Deepwater Gulf of Mexico}: Society of Exploration Geophysicists.

\bibitem[Gao et~al., 2021]{gao2020threedimensionalseismiccharacterizationimaging}
Gao, K., L. Huang, and T. Cladouhos,  2021, {Three-dimensional seismic characterization and imaging of the Soda Lake geothermal field}: Geothermics, {\bfseries 90}, 101996.

\bibitem[Gao et~al., 2024]{Gao2024MultiscaleResidualConvolution}
Gao, Z., H. Chen, Z. Li, and B. Ma,  2024, {Multiscale Residual Convolution Neural Network for Seismic Data Denoising}: IEEE Geoscience and Remote Sensing Letters, {\bfseries 21}, 1--5.

\bibitem[Gatys et~al., 2016]{neural_style_transfer}
Gatys, L.~A., A.~S. Ecker, and M. Bethge,  2016, {Image Style Transfer Using Convolutional Neural Networks}: 2016 IEEE Conference on Computer Vision and Pattern Recognition (CVPR), 2414--2423.

\bibitem[George et~al., 2018]{George2018Surveymedian}
George, G., R.~M. Oommen, S. Shelly, S.~S. Philipose, and A.~M. Varghese,  2018, {A Survey on Various Median Filtering Techniques For Removal of Impulse Noise From Digital Image}: 2018 Conference on Emerging Devices and Smart Systems (ICEDSS), 235--238.

\bibitem[Goyes-Pe{\~n}afiel et~al., 2024]{goyespenafiel2024gansupervisedseismicdatareconstruction}
Goyes-Pe{\~n}afiel, P., L. Su{\'a}rez-Rodr{\'\i}guez, C.~V. Correa, and H. Arguello,  2024, {GAN--supervised Seismic Data Reconstruction: An Enhanced--Learning for Improved Generalization}: IEEE Transactions on Geoscience and Remote Sensing.

\bibitem[Hlebnikov et~al., 2021]{Volodya2021Noisetypes}
Hlebnikov, V., T. Elboth, V. Vinje, and L.-J. Gelius,  2021, {Noise types and their attenuation in towed marine seismic: A tutorial}: GEOPHYSICS, {\bfseries 86}, W1--W19.

\bibitem[Holbrook, 2015]{blakeridge}
Holbrook, S.,  2015, {Project Blake Ridge Hydrates 3{D}}.

\bibitem[Howard et~al., 2018]{tgs}
Howard, A., A. Sharma, A. Lenamond, J. Adamck, M. McDonald, S. Kainkaryam, and W. Cukierski,  2018, {TGS Salt Identification Challenge}.

\bibitem[Jia et~al., 2024]{jia2024groundrollseparationlandseismic}
Jia, Z., W. Lu, M. Zhang, and Y. Miao,  2024, {Ground-roll Separation From Land Seismic Records Based on Convolutional Neural Network}.

\bibitem[Karras et~al., 2018]{karras2018progressivegrowinggans}
Karras, T., T. Aila, S. Laine, and J. Lehtinen,  2018, {Progressive Growing of {GAN}s for Improved Quality, Stability, and Variation}: Presented at the International Conference on Learning Representations.

\bibitem[Keys and Foster, 1994]{mobil_avo_1994}
Keys, R.~G., and D.~J. Foster, eds., 1994, Mobil avo viking graben line 12 data released for 1994 seg workshop: Society of Exploration Geophysicists (SEG).
\newblock SEG File Publications, No.~No 4.

\bibitem[Kingma and Ba, 2015]{kingma2017adammethodstochasticoptimization}
Kingma, D., and J. Ba,  2015, Adam: A method for stochastic optimization: Presented at the {International Conference on Learning Representations (ICLR)}.

\bibitem[Kumar and Ahmed, 2021]{Kumar2021Seismic}
Kumar, D., and I. Ahmed,  2021, 1, {\itshape in} {Seismic Noise}: Springer International Publishing,  1442--1447.

\bibitem[Larsen et~al., 2016]{larsen2016autoencoding}
Larsen, A. B.~L., S.~K. S{\o}nderby, H. Larochelle, and O. Winther,  2016, Autoencoding beyond pixels using a learned similarity metric: International conference on machine learning, PMLR, 1558--1566.

\bibitem[Liu et~al., 2024]{Liu2024SeismicRandomNoiseSuppression}
Liu, X., F. Lyu, L. Chen, C. Li, S. Zu, and B. Wang,  2024, {Seismic Random Noise Suppression Based on Deep Image Prior and Total Variation}: IEEE Transactions on Geoscience and Remote Sensing, {\bfseries 62}, 1--11.

\bibitem[Malehmir et~al., 2021]{Malehmir2021Sparse3Dreflectionseismicsurvey}
Malehmir, A., M. Markovic, P. Marsden, A. Gil, S. Buske, L. Sito, E. B\"ackstr\"om, M. Sadeghi, and S. Luth,  2021, {Sparse 3D reflection seismic survey for deep-targeting iron oxide deposits and their host rocks, Ludvika Mines, Sweden}: Solid Earth, {\bfseries 12}, 483--502.

\bibitem[Martin et~al., 2006]{AGL}
Martin, G.~S., R. Wiley, and K.~J. Marfurt,  2006, Marmousi2: An elastic upgrade for marmousi: The Leading Edge, {\bfseries 25}, 156--166.

\bibitem[Minerals, 1995]{kerry}
Minerals, N. Z.~C.,  1995, {Kerry 3{D} dataset}.

\bibitem[Mrigya et~al., 2023]{mrigya2023convolutional}
Mrigya, F., R. Samiran, F. Viviane, and H. Satyan~Singh,  2023, {A Comparative Analysis of Convolutional Neural Networks for Seismic Noise Attenuation}: SPE EuropEC - Europe Energy Conference featured at the 84th EAGE Annual Conference \& Exhibition.

\bibitem[Oktay et~al., 2018]{oktay2018attentionunetlearninglook}
Oktay, O., J. Schlemper, L.~L. Folgoc, M. Lee, M. Heinrich, K. Misawa, K. Mori, S. McDonagh, N.~Y. Hammerla, B. Kainz, B. Glocker, and D. Rueckert,  2018, {Attention U-Net: Learning Where to Look for the Pancreas}: Presented at the Medical Imaging with Deep Learning.

\bibitem[Oumarou et~al., 2021]{Oumarou2021}
Oumarou, S., D. Mabrouk, T.~C. Tabod, J. Marcel, S.~N. III, J.~M.~A. Essi, and J. Kamguia,  2021, {Seismic attributes in reservoir characterization: an overview}: Arabian Journal of Geosciences, {\bfseries 14}, 402.

\bibitem[Persson, 2021]{progan}
Persson, A.,  2021, {ProGAN implementation in PyTorch}: \url{https://github.com/aladdinpersson/Machine-Learning-Collection/tree/master/ML/Pytorch/GANs/ProGAN}.
\newblock (Accessed: 2024-11-16).

\bibitem[Qian et~al., 2024]{Qian2024Unsupervised}
Qian, F., H. Hua, Y. Wen, S. Pan, G. Zhang, and G. Hu,  2024, {Unsupervised 3-D Seismic Erratic Noise Attenuation With Robust Tensor Deep Learning}: IEEE Transactions on Geoscience and Remote Sensing, {\bfseries 62}, 1--16.

\bibitem[Ren et~al., 2022]{Hongping2022Denoising}
Ren, H., X. Wen, and C. Tang,  2022, {Denoising seismic data with drilling noise based on GAN}: {SEG 2021 Workshop: 4th International Workshop on Mathematical Geophysics: Traditional {\&} Learning, Virtual, 17–19 December 2021}, Society of Exploration Geophysicists, 126--128.

\bibitem[Sam and Kurt, 1995]{1994bp}
Sam, G., and M. Kurt,  1995, {Migration from topography: Improving the near-surface image}.

\bibitem[Ulyanov et~al., 2020]{Ulyanov_2020}
Ulyanov, D., A. Vedaldi, and V. Lempitsky,  2020, {Deep Image Prior}: International Journal of Computer Vision, {\bfseries 128}, 1867–1888.

\bibitem[Van Den~Oord et~al., 2017]{van2017neural}
Van Den~Oord, A., O. Vinyals, et~al.,  2017, Neural discrete representation learning: Advances in neural information processing systems, {\bfseries 30}.

\bibitem[Wang et~al., 2024]{Wang2024Self-Supervised}
Wang, H., J. Lin, Y. Li, X. Dong, X. Tong, and S. Lu,  2024, {Self-Supervised Pretraining Transformer for Seismic Data Denoising}: IEEE Transactions on Geoscience and Remote Sensing, {\bfseries 62}, 1--25.

\bibitem[Wang et~al., 2025]{Wang2024QuadraticUnet}
Wang, X., Y. Sui, and J. Ma,  2025, {Quadratic Unet for seismic random noise attenuation}: Geophysics, {\bfseries 90}, V43--V55.

\bibitem[Wei et~al., 2022]{WEI2022104968}
Wei, X.-L., C.-X. Zhang, S.-W. Kim, K.-L. Jing, Y.-J. Wang, S. Xu, and Z.-Z. Xie,  2022, {Seismic fault detection using convolutional neural networks with focal loss}: Computers \& Geosciences, {\bfseries 158}, 104968.

\bibitem[Wu et~al., 2022]{WU2022110431}
Wu, H., B. Zhang, and N. Liu,  2022, {Self-adaptive denoising net: Self-supervised learning for seismic migration artifacts and random noise attenuation}: Journal of Petroleum Science and Engineering, {\bfseries 214}, 110431.

\bibitem[Xu et~al., 2023]{Xu2023s2s-wtv}
Xu, Z., Y. Luo, B. Wu, and D. Meng,  2023, {S2S-WTV: Seismic Data Noise Attenuation Using Weighted Total Variation Regularized Self-Supervised Learning}: IEEE Transactions on Geoscience and Remote Sensing, {\bfseries 61}, 1--15.

\bibitem[Yang et~al., 2023]{Yang2023randomnoiseattenuationDnCNN}
Yang, W., X. Chen, and Y. Rao,  2023, {{Seismic random noise attenuation using DnCNN with stratigraphic dip constraint}}: Journal of Geophysics and Engineering, {\bfseries 20}, 1172--1179.

\bibitem[Yoo and Zwartjes, 2022]{YOO2022123}
Yoo, J., and P. Zwartjes,  2022, {Attenuation of seismic migration smile artifacts with deep learning}: Artificial Intelligence in Geosciences, {\bfseries 3}, 123--131.

\bibitem[Zhang et~al., 2024]{Zhang2024}
Zhang, Y., C. Zhang, and L. Song,  2024, {A two-stage seismic data denoising network based on deep learning}: Studia Geophysica et Geodaetica.

\end{thebibliography}

\end{document}